\documentclass[twocolumn,aps,prl,floats,amsfonts]{revtex4}
\usepackage{graphicx}

\newcommand {\be}{\begin{equation}}
\newcommand {\ee} {\end{equation}}
\newcommand {\ba}{\begin{eqnarray}}
\newcommand {\ea} {\end{eqnarray}}

\begin{document}


\title{Non-Abelian quantized Hall states of electrons at filling 
factors $12/5$ and
$13/5$\\
in the first excited Landau level}
\author{E.H. Rezayi$^{1}$ and N. Read$^{2}$}
\address{$^1$Department of Physics, California State University,
Los Angeles, CA 90032}
\address{$^2$Department of Physics, Yale University, P.O. Box
208120, New Haven, CT 06520-8120}

\date{\today}

\begin{abstract}
We present results of extensive numerical calculations on the
ground state of electrons in the first excited ($n=1$) Landau
level with Coulomb interactions, and including non-zero thickness
effects, for filling factors $12/5$ and $13/5$ in the torus
geometry. In a region that includes these experimentally-relevant
values, we find that the energy spectrum and the overlaps with the
trial states support the previous hypothesis that the system is in
the non-Abelian $k=3$ liquid phase we introduced in a previous
paper.
\end{abstract}

\maketitle

Many distinct quantum Hall liquid phases have been observed in
two-dimensional electron systems. Leaving aside those that occur
at integer filling factor (or quantized Hall conductance), the
non-integer fractions occur at low filling factors in high
mobility samples (at higher filling factors, they are supplanted
by states in which translational or rotational symmetry is
violated, or by ``re-entrant'' integer quantized Hall phases). In
the lowest ($n=0$) Landau level (LL), that is at filling factors
less than $2$, the incompressible liquids are phases of matter
that are correctly characterized by the Laughlin states
\cite{laugh}, and their extensions via the hierarchy \cite{hier}
or composite fermion \cite{jain} approaches; these two approaches
describe the same phases \cite{rbw}. In the first excited ($n=1$)
Landau level, the physics appears to be different, as one gets
closer to the broken symmetry phases. In general, fractions in the
$n=1$ LL can be compared with those in the $n=0$ level by
subtracting $2$ from the filling factor; this corresponds to
simply filling the lowest level with both spins, and treating the
next level like the lowest. A quantized Hall plateau occurs at
fillings $5/2$ and $7/2$, which is now widely believed
\cite{morf,rh} to correspond to an incompressible liquid that may
be viewed as a p-wave paired state of spin-polarized composite
fermions in a half-filled LL \cite{mr}. It is unlike the lowest
Landau level (LLL) case, in which at filling factor $\nu=1/2$ or
$3/2$ the composite fermion liquid appears so far to be gapless,
and exhibits Fermi-liquid--like properties.

A few other fractions are observed between $2$ and $4$, and one
may wonder whether these are the same phases as occur in the LLL.
In an earlier paper \cite{rr99} (to be referred to as RR), a
sequence of incompressible fractional quantum Hall liquids was
constructed. These have filling factors $\nu=k/(Mk+2)$, where
$k=1$, $2$, $3$, \ldots, and $M=0$, $1$, $2$, are integers, and
for fermions $M$ must be odd (the values $[(M-1)k+2]/(Mk+2)$ 
may
also be obtained, by using particle-hole symmetry). These were
constructed within the LLL, but may be applied to higher filling
factors by adding the filling of the lower levels. The $k=1$
liquids are the familiar Laughlin states \cite{laugh}, while $k=2$
is the Moore-Read (MR) paired state \cite{mr}. For the next case,
$k=3$, some evidence that it occurs in the $n=1$ Landau level
(with $M=1$, so $\nu=2+\frac{3}{5}$ or $2+\frac{2}{5}$) was
presented, but was perhaps not entirely convincing. Meanwhile,
experiments have observed a fractional quantum Hall state at
$12/5$, which exhibits a remarkably small energy gap for charged
excitations \cite{xia}. There is a great deal of interest in
phases like the RR states, as for $k>1$ they exhibit excitations
that obey non-Abelian statistics \cite{mr}, and for $k\neq 1$, $2$
or $4$ they would support universal quantum computation
\cite{freedman}.

In this paper we return to the issue of the nature of the liquid
state at filling fractions $12/5$ and $13/5$. We report numerical
calculations for moderate numbers of electrons in a single LL,
with interactions that model the Coulomb interaction between
electrons in the $n=1$ LL, plus the effect of non-zero thickness
of the electron wavefunction in the direction perpendicular to the
two-dimensional layer. We also explore the phase diagram as the
short-range component of the interaction is varied. For
experimentally relevant values of thickness, we find that the
liquid phase in the vicinity of the $n=1$ Coulomb interaction
appears to be the RR $k=3$ phase. The evidence for this comes 
from
the spectrum on the torus (i.e.\ periodic boundary conditions on a
parallelogram), which exhibits a doublet of ground states that is
characteristic of the $k=3$ phase, separated from higher excited
states by a significant gap; this doublet does not occur for the
hierarchy/composite-fermion (H-CF) phase. Quantitatively, the gap
in the spectrum is small, suggesting that the gap for charged
excitations will also be much smaller than for the H-CF 2/5 state
in the LLL, in general agreement with experiment. The states in
the ground state doublet have large overlaps with the trial states
of RR, adapted to the torus. When the interaction is modified,
phase transitions to a ``stripe'' phase, and to the usual H-CF
phase are observed. Several of these results, except possibly for
this last transition, are similar to results of a recent study
\cite{rrc} of bosons in the LLL at filling $3/2$, which is the
$k=3$, $M=0$ case of RR, and also to the $1/2$ (or $5/2$, etc)
case for electrons \cite{rh}.

The methods of calculation are rather standard, so we will be
brief. For spin-polarized electrons in two dimensions confined to
any one LL, the interaction Hamiltonian $H_{\rm int}$ in the
infinite plane can be represented by the pseudopotentials $V_m$,
$m=1$, $3$, \ldots. $V_m$ is the interaction energy for a single
pair of electrons of relative angular momentum $m$ ($m$ odd)
\cite{hald83}. For the case of zero thickness, the
pseudopotentials for the Coulomb interaction in the $n=1$ LL were
plotted in Ref.\ \cite{haldbook}. We also include non-zero
thickness through the standard Fang-Howard method, in which the
wavefunction contains the dependence $\propto x_3 e^{-x_3/(2b)}$
(but vanishes for $x_3<0$) on the perpendicular coordinate $x_3$.
This tends to reduce somewhat the low $m$ pseudopotentials
relative to the others. We can then represent any one LL using the
states in the LLL. This description of the interaction can be
extended to the sphere (also using rotation symmetry), and to the
torus; in these cases there are $N_\phi$ flux quanta piercing the
system. Starting from the pseudopotentials for the non-zero
thickness $n=1$ LL, we also consider changing the first two
pseudopotentials $V_1$, $V_3$ by small amounts $\delta V_1$,
$\delta V_3$ in order to explore the phase diagram in the vicinity
of this interaction. This mitigates to some extent our ignorance
of the precise interaction Hamiltonian. For comparison, for $15$
particles on the torus and $w=2\ell_B$, the unperturbed values are
$V_1=0.3858$, $V_3=0.3333$. We also note that particle-hole
symmetry holds as long as inter-LL interactions are neglected, as
here, so that our results for $\nu=13/5$ also apply to $12/5$,
with no modifications at all in the case of the torus geometry.
Likewise, our results should also be relevant for filling factors
$3+\frac{2}{5}$ and $3+\frac{3}{5}$, which also lie in the $n=1$
LL. Accordingly, we refer only to $\nu=3/5$ from here on.

We also refer to a (positive) $p$-body interaction %
which penalizes the closest approach of $p$ fermions that is
allowed by Fermi statistics; it can be written in terms of
derivatives of $\delta$-functions on the sphere or torus. For
fermions (electrons), the parafermion trial states found in ref.\
\cite{rr99} are unique, exact zero-energy eigenstates of such
interactions for $p=k+1$ when $N$ is divisible by $k$ and
$N_\phi=(k+2)N/k -3$ (on the sphere), so that
$\nu=\lim_{N\to\infty} N/N_V=k/(k+2)$ (thus $M=1$). There are
corresponding states on the torus, for $N_\phi=(k+2)N/k$. For the
$k+1$-body interaction, the trial states represent incompressible
liquid phases, in which the excitations enjoy non-Abelian
statistics for $k>1$. The $k+1$-body Hamiltonians allow us to
numerically generate the trial states, for comparison with the
exact ground states of the two-body pseudopotential interaction at
the same $N$, $N_\phi$ in the same geometry.

On the torus at $\nu=N/N_\phi=3/5$, translational symmetry 
implies
that all energy eigenstates possess a trivial center-of-mass
degeneracy of $5$, which is exact for any size system, and also
that there is a quantum number called $\bf K$ \cite{hald85}, which
is a vector lying in a certain Brillouin zone. For ${\bf K=0}$,
energy eigenstates can also be labeled by their eigenvalues under
a rotation. In the $M=1$ RR phases, the ground states have a net
degeneracy $\frac{1}{2}(k+1)(k+2)$ in the thermodynamic limit,
which is connected with the non-Abelian statistics \cite{mr,rr99}.
For $k=3$, this 10-fold degeneracy is made up of the trivial
factor $5$ (which is always discarded in numerical studies),
together with a further $2$-fold degeneracy, which in general
becomes exact only in the thermodynamic limit; all these ground
states have ${\bf K=0}$. (For the $4$-body interaction, the
10-fold degeneracy is exact for any size, as the ground states
have exactly zero energy.) By contrast, the H-CF ground state for
fermions at $\nu=3/5$ possesses only the $5$-fold degeneracy. 
Then
for an incompressible fluid on the torus, the spectrum of a
sufficiently large system in one of these two phases should
exhibit a nearly degenerate pair of ground-states or a single
ground state, respectively, at $\bf K = 0$, separated by a clear
gap from a region of many states at higher energy eigenvalues.

\begin{figure}
{\centering \includegraphics[width=3.0in]{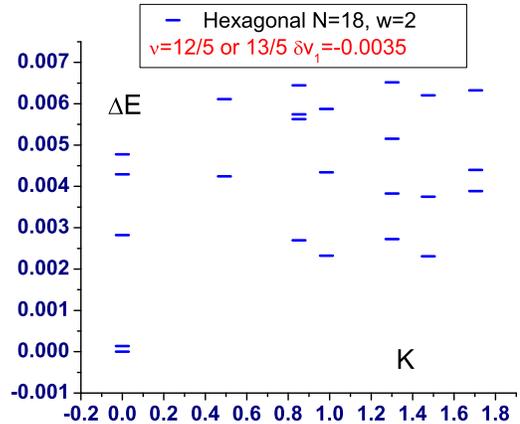} }
\caption{ \label{fig:specpbc} Low-lying spectrum for $18$
electrons on the torus on the torus vs.\ $K=|{\bf K}|$, for
hexagonal unit cell. Energies are in units of $e^2/\ell_B$. The
interaction Hamiltonian is the $n=1$ Coulomb interaction,
including non-zero thickness with parameter $w=2\ell_B$, and
$\delta V_1=-0.0035$.}
\end{figure}

In Fig.\ \ref{fig:specpbc} we show the spectrum for $18$ electrons
on the torus at $30$ flux quanta, with thickness parameter
$w=2b=2$, and a small change to the pseudopotentials, $\delta
V_1=-0.0035$, $\delta V_3=0$. In this and all subsequent figures,
the geometry of the torus is the hexagonal unit cell. Energies are
in units of $e^2/\ell_B$ for the Coulomb interaction $e^2/r$. We
set the magnetic length $\ell_B$ and $\hbar$ to one throughout.
For this case, a ground-state doublet at ${\bf K=0}$ is apparent,
separated from the excited states by a gap which is about ten
times larger than the splitting of the doublet, or than the
typical spacings above the gap. This is the expected behavior of
the RR incompressible fluid phase. However, we will discuss below
the variation of this splitting with the interaction parameters.

\begin{figure} 
{\centering \includegraphics[width=3.0in]{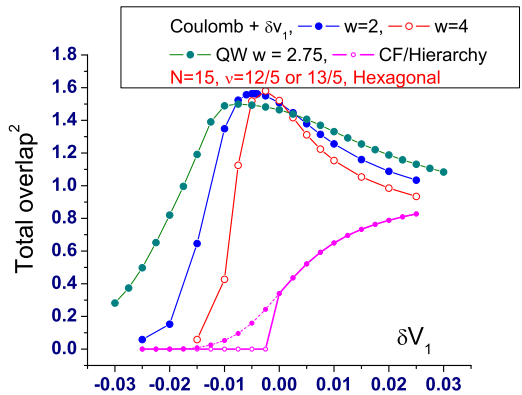} }
\caption{ \label{fig:overlaps} The sum of the squared-overlap of
each member of the ground-state doublet for $n=1$ Coulomb
interaction (for $w=2$ and $w$) with the trial two-dimensional
subspace, as a function of $\delta V_1$ on the torus, for $18$
particles. [Also included is the same for a quantum well
wavefunction in $x_3$, labeled QW, with $w=2.75$.] Also, the
squared-overlap of one of the ground state doublet for $w=2$ with
the H-CF state as a function of $\delta V_1$ in the same system:
solid line--lowest in doublet; dashed line--next lowest.}
\end{figure}

In Fig.\ \ref{fig:overlaps}, we show the sum of the
squared-overlaps of the lowest two ${\bf K=0}$ states with the
low-lying doublet in Fig.\ \ref{fig:specpbc} with the
two-dimensional subspace of zero-energy states of $H_4$ (the trial
states) for $N=15$ particles, as a function of $\delta V_1$
($\delta V_3=0$), for two values of $w=2b$. The same is also 
shown
for another model of non-zero thickness effect, which describes a
quantum well in the $x_3$ direction, as used in some experimental
samples; for this the thickness is $w=2.75$. In addition, we have
plotted the squared-overlap of each of the two states in the
doublet with the single trial ground state for the H-CF phase, for
$w=2$. For the latter, we have used the numerically-obtained
ground state for a very large positive value of $\delta V_1$. In
that regime, which is similar to the LLL Coulomb pseudopotential
values, the ground state is believed to be incompressible and to
lie in the H-CF phase. The H-CF state has exactly zero overlap
with one member of the doublet, while the overlap of the other
with the H-CF state is not zero. The overlap of the H-CF state
with the lowest ${\bf K=0}$ state drops abruptly to zero for
$\delta V_1$ less than about $0$, which is due to a level
crossing. Beyond that point, the other ${\bf K=0}$ state is lower,
and has zero overlap with the H-CF state, but the overlap of the
other with the H-CF state remains non-zero and is shown as the
dashed line. This behavior, both the vanishing overlap and the
level crossing, indicates that the two members of the doublet have
different rotational symmetry in the hexagonal geometry, and only
one has the same symmetry as the H-CF ground state. The non-
zero
overlap with the H-CF state is much smaller than that with the
two-dimensional RR trial subspace, but increases with increasing
$\delta V_1$, while the overlaps with RR slowly decrease (this
behavior is consistent with that observed on the sphere at smaller
sizes \cite{rr99}; note that on the sphere the RR and H-CF states
occur at different $N_\phi$, which is not the case on the torus).
These results suggest that there must be a phase transition in the
thermodynamic limit between these two incompressible phases at
some value of $\delta V_1$, and this occurs for a range of values
of thickness $w$.

\begin{figure} 
{\centering \includegraphics[width=3.0in]{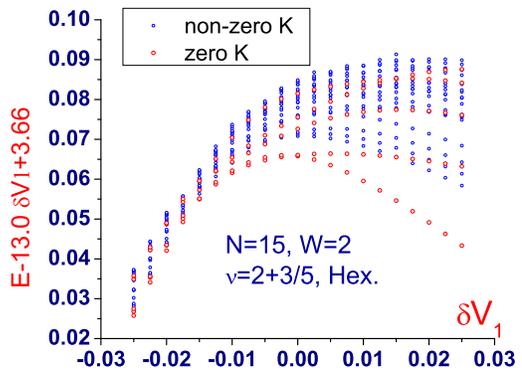} }
\caption{ \label{fig:specvsv1} The low-lying energy spectrum as a
function of $\delta V_1$ on the torus, for $15$ particles. The
${\bf K=0}$ and ${\bf K} \neq {\bf 0}$ levels are shown as
distinct symbols.}
\end{figure}

In Fig.\ \ref{fig:specvsv1}, we show the low-lying energy spectrum
for $N=15$ particles as the $\delta V_1$ is varied away zero, for
$w=2$, and $\delta V_3=0$. The level-crossing of the lowest two
${\bf K=0}$ levels, mentioned in the previous paragraph, can be
seen at $\delta V_1$ close to $0$. Over the approximate range
$-0.01 < \delta V_1 <0$, the lowest two ${\bf K=0}$ states stay
remarkably close in energy, and separated from all other states in
the system. For $\delta V_1$ negative, a transition occurs at
around $-0.01$ (a similar transition was seen in Refs.\
\cite{rh,rrc}). Beyond that point, the spectrum as a function of
aspect ratio of the torus (not shown) shows signs that the system
is in a stripe phase. Fig.\ \ref{fig:specvsv1} reveals that this
transition is the strongest feature in the spectrum, which
dominates the behavior of moderately low-lying levels even rather
far away in $\delta V_1$. The behavior of the spectrum, with many
levels converging to zero at the same point, suggests that this
transition may be second-order in the thermodynamic limit. In
addition, it indicates that the energy scales in the RR phase are
smaller than in the H-CF phase which occurs at large positive
$\delta V_1$. The transition between the latter two phases (where
the higher ${\bf K=0}$ state moves up into the continuum) occurs
not far from the level crossing of ground states. In contrast to
the rapid rise in energy of one of the two ground states of the RR
region on entering the H-CF region, the higher energy levels show
only gradual changes in this range of $\delta V_1$. This might
suggest that this transition is first-order in the thermodynamic
limit, but see also the spectra for $18$ particles below.

\begin{figure}  
{\centering \includegraphics[width=3.0in]{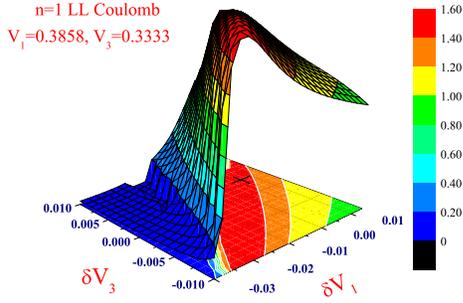} }
\caption{ \label{fig:3doverlaps} A 3-dimensional plot of the total
squared-overlap with the RR trial subspace for $N=15$ particles on
the torus, with $w=2$, as a function of both $\delta V_1$ and
$\delta V_3$.}
\end{figure}

\begin{figure}  
{\centering \includegraphics[width=3.0in]{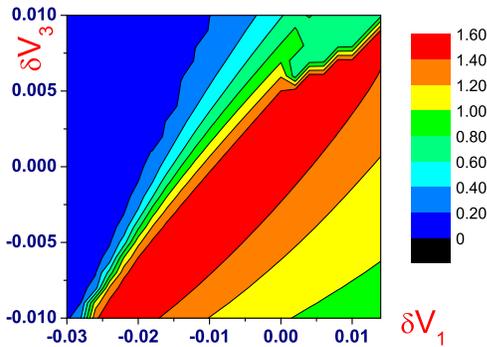} }
\caption{ \label{fig:contover} The same as Fig.\
\protect{\ref{fig:3doverlaps}}, but shown as a contour plot.}
\end{figure}

In Fig.\ \ref{fig:3doverlaps} we show a 3-dimensional plot of the
total squared-overlap with the RR trial subspace for $N=15$
particles on the torus, with $w=2$, as a function of both $\delta
V_1$ and $\delta V_3$. This shows that the high, broad maximum 
as
a function of $\delta V_1$ persists over a range of $\delta V_3$,
though the location shifts.  The same information is shown in
Fig.\ \ref{fig:contover} as a contour plot. We note that the
squared-overlap falls rapidly towards zero as the stripe region is
entered. In view of the large dimensions of the Hilbert spaces of
${\bf K=0}$ states, the largest squared-overlaps ($>1.40$) should
be viewed as significantly large.

\begin{figure}   
{\centering \includegraphics[width=3.0in]{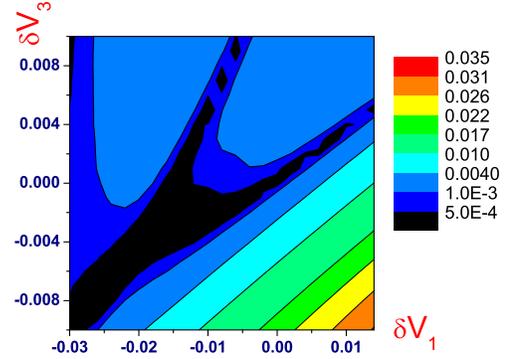} }
\caption{ \label{fig:contgap} A contour plot of the level
splitting between the two lowest ${\bf K=0}$ states for $15$
particles as a function of $\delta V_1$ and $\delta V_3$.}
\end{figure}

In Fig.\ \ref{fig:contgap}, we show the splitting between the two
lowest ${\bf K=0}$ states for $15$ particles as a function of
$\delta V_1$ and $\delta V_3$, also for $w=2$. For $\delta V_3$
zero or negative, the splitting is smallest close to where the
overlap with the RR states is largest, however at larger $\delta
V_3$ the region of smallest splitting is seen to bifurcate, while
the overlaps there are unaffected.

\begin{figure}  
{\centering \includegraphics[width=3.0in]{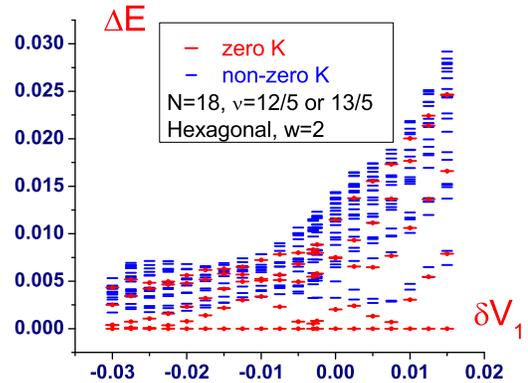} }
\caption{ \label{fig:18specvsv1} Low-lying energy spectrum, shown
as the difference from the ground state energy, for $18$ particles
as a function of $\delta V_1$. Again, states at ${\bf K=0}$ and
${\bf K} \neq {\bf 0}$ are shown with distinct symbols.}
\end{figure}

Finally, we show in Fig.\ \ref{fig:18specvsv1} the low-lying
spectrum for $N=18$ particles as a function of $\delta V_1$, for
$\delta V_3=0$. Overall features are similar to those for $15$
particles in Fig.\ \ref{fig:specvsv1}, however, here the splitting
of the lowest two ${\bf K=0}$ states in the possible RR region is
not as small as for $N=15$, and there are now two crossings
between these two states between the strip region and the H-CF
region (the spectrum in Fig.\ \ref{fig:specpbc} is taken near one
of these crossings). In addition, the gap to the non-zero $\bf K$
levels decreases somewhat in the transition region to the H-CF,
and is even comparable to the splitting of the doublet in that
region (however the gap to other ${\bf K=0}$ states remains
larger). Nonetheless, we expect that overlaps with the RR states
remain large, and those with the H-CF state, small, in the central
region with a strong gap above the ground state. We note that the
appearance of two level crossings, with a similar not-so-small
splitting between them also seems to occur in the $N=15$ case if
we examine larger values of $\delta V_3$ (the ``bifurcation''
noted above), while overlaps with the RR subspace remain large.
Consequently, this aspect of the $N=18$ data seems consistent 
with
behavior at $N=15$.

Uniting all aspects of our results, we conclude that over a range
of parameter values that includes the $n=1$ Coulomb interaction
and experimentally relevant non-zero thickness effects, there is
large overlap of the lowest two $\bf K=0$ states with the trial
subspace, and relatively small splittings of those two states,
suggesting that they become degenerate in the thermodynamic 
limit.
This region is bordered on one side by the H-CF phase, which has
poor overlap (and for some parameters {\em zero} overlap, due to
different symmetry) with the lowest member of the doublet in this region. It is
bordered on the other side by a stripe phase, similar to others in
the higher LLs. The overall pattern of behavior is quite similar
to that observed for the MR phase at $5/2$ \cite{rh}, and is also
consistent with our previous results on the present system on the
sphere \cite{rr99}.

To conclude, we find the evidence that the observed 12/5 state is
in the RR phase compelling. The non-observation of the $13/5$
state so far suggests that LL mixing plays a role at the latter
filling factor. While a systematic study of energy gaps for
charged excitations will have to await larger system sizes, we
find indications that the scales in this state will be smaller
than in the H-CF phase at the same density of particles, which is
broadly consistent with experimental findings \cite{xia}.

We thank N.R. Cooper for stimulating discussions. EHR thanks D.
Haldane for use of his code in some of the calculations. This work was supported by DOE
contract no.\ DE-FG03-02ER-45981 (EHR), and initially by NSF 
grant
DMR0086191 (EHR), as well as by NSF grant no.\ DMR-02-42949 
(NR).



\begin{references}



\bibitem{laugh}
   R.B. Laughlin, Phys.\ Rev.\ Lett.\ {\bf 50}, 1395 (1983).

\bibitem{hier}
   F.D.M. Haldane, Phys.\ Rev.\ Lett.\ {\bf 51}, 605 (1983);
   B.I. Halperin, Phys.\ Rev.\ Lett.\ {\bf 52}, 1583 (1984).

\bibitem{jain} J.K. Jain, Phys. Rev. Lett. {\bf 63}, 199 (1989).

\bibitem{rbw} N. Read, Phys. Rev Lett. {\bf 65}, 1502 (1990);
B. Blok and X.-G. Wen, Phys Rev. B {\bf 42}, 8133, 8145 (1990);
{\it ibid.}, {\bf 43}, 8337 (1991).

\bibitem{morf} R.H. Morf, Phys. Rev. Lett. {\bf 80}, 1505 (1998).

\bibitem{rh} E.H. Rezayi and F.D.M. Haldane, Phys. Rev. Lett.
{\bf 84}, 4685 (2000).

\bibitem{mr}
G.~Moore and N.~Read, Nucl.\ Phys.\ {\bf B 360}, 362 (1991).

\bibitem{rr99}
N. Read and E. Rezayi, \prb {\bf 59}, 8084 (1999).

\bibitem{xia} J.S. Xia {\it et al.}, Phys. Rev. Lett. { 93},
176809 (2004).

\bibitem{freedman} M.H. Freedman, A. Kitaev, M.J. Larsen, and
Z. Wang, quant-ph/0101025.

\bibitem{rrc} E.H. Rezayi, N. Read, and N.R. Cooper,
Phys. Rev. Lett. {\bf 95}, 160404 (2005).

\bibitem{hald83}
F.D.M. Haldane, in Ref.\ \cite{hier}.

\bibitem{haldbook} F.D.M. Haldane, Fig.\ 8.1 in {\it The Quantum 
Hall Effect},
  edited by R.E.~Prange and S.M.~Girvin (Second Edition, Springer-
Verlag,
  New York, 1990).

\bibitem{hald85}
F.D.M. Haldane, Phys. Rev. Lett. {\bf 55}, 2095 (1985).

\end{references}
\end{document}